\documentclass{article}

\usepackage{graphicx}
\usepackage{amssymb}

\usepackage{lineno}

\usepackage{longtable}
\usepackage{booktabs}
\usepackage{amsmath}
\usepackage{algorithm}
\usepackage[noend]{algpseudocode}
\usepackage{siunitx}
\usepackage{subcaption}
\usepackage[hyphens]{url}
\usepackage{natbib}
\usepackage{todonotes}
\usepackage{color}
\usepackage{float}
\usepackage{authblk}

\begin{document}

\title{The Sensitivity of Electric Power Infrastructure Resilience to the Spatial Distribution of Disaster Impacts}

\author[$\dagger$]{Rachunok, Benjamin}
\author[$\dagger \star$]{Nateghi, Roshanak}

\affil[$\dagger$]{School of Industrial Engineering, Purdue University, West Lafayette, IN}
\affil[$\star$]{Division of Environmental and Ecological Engineering, Purdue University, West Lafayette, IN}
\maketitle
\begin{abstract}
Credibly assessing the resilience of energy infrastructure in the face of natural disasters is a salient concern facing researchers, government officials, and community members. Here, we explore the influence of the spatial distribution of disruptions due to hurricanes and other natural hazards on the resilience of power distribution systems. We find that incorporating information about the spatial distribution of disaster impacts has significant implications for estimating infrastructure resilience. Specifically, the uncertainty associated with estimated infrastructure resilience metrics to spatially distributed disaster-induced disruptions is much higher than determined by previous methods. We present a case study of an electric power distribution grid impacted by a major landfalling hurricane. We show that improved characterizations of disaster disruption drastically change the way in which the grid recovers, including changes in emergent system properties such as antifragility. Our work demonstrates that previous methods for estimating critical infrastructure resilience may be overstating the confidence associated with estimated network recoveries due to the lack of consideration of the spatial structure of disruptions. 

\end{abstract}


\section{Introduction}

Defined broadly, resilience is an emergent property of a system which manifests as the result of an iterative process of sensing, anticipation, learning, and adaptation to all types of disruptions \cite{park2013}. Using this definition, resilience must be studied at a system-wide level, where the resilience of an entire system is studied in the context of hazards and disruptions. Characterization of the resilience of a complex system, therefore, is inherently a comprehensive analysis of that which acts against it. This system--disruption paradigm allows for the study of a wide range of interaction-based entities from ecological plant--pollinator relationships \cite{kaiser2017,holling1973} to the psychological resilience of families to trauma \cite{riggs2011}. 

In the context of engineering urban systems, the resilience of a critical infrastructure (e.g., the electric power grid, telecommunication networks, natural gas, water network, etc.,) \emph{includes} study of the recovery from failures induced by hydro-climatic extremes and seismic events as well as acts of terrorism. Critical urban networked infrastructure is well-represented  by a graph \cite{newman2018}. Subsequently, disrupting a graph requires removing or disabling fractions of the system consistent with an exogenous threat or hazard.

In this paper, we use a graph-theoretic approach to show that small changes in the spatial characteristics of a disruption to a system radically change the characteristics of system performance as a disruption is repaired over time. Whether the recovery is measured in-terms of network-based performance metrics or by the extent of impact on stakeholders, our results indicate that the measured resilience of a system is heavily dependant on the spatial characteristics of the initial disruption. We conduct this study in the case of an electric power distribution grid impacted by a major landfalling hurricane. We generate different \emph{spatial} distributions of initial disruptions to a power grid and study their impact on graph-theoretic measures of network connectivity as well as the number of customers without power. The remainder of this paper is as follows: Section 2 introduces relavent other works, Section 3 outlines the data and methods used for this analysis, and finally Sections 4 and 5 detail the results and conclusion respectively.

\section{Background}

Network analysis 
deals with the study of graphs or networks. Networks are ``a collection of points [referred to as \emph{vertices} or \emph{nodes}] joined together by pairs of lines [referred to as  \emph{edges} or \emph{links}].'' \cite{newman2018} The edge-vertex pairing lends itself to be an intuitive mathematical object for which to model phenomenon such as animal and plant interactions \cite{dietze2018}, academic authorship, urban infrastructure design \cite{derrible2017} \cite{clauset2009} and---most relevant to this work---electric power infrastructure \cite{nan2017,hines2010,larocca,duenas-osorioleonardo2007}. Representing a system as a network allows for simple---and in most cases tractable---estimations of system performance. Measurements of the overall size, degree of connectivity, length of paths between vertices, and degree of clustering are all easily computed from a network model and can provide a myriad of insights about the system being represented \cite{barabasi2016}. Graphs representing a system in which the components interact can be used to model how the failure of one vertex may propagate through the network \cite{hu2016}. If failure likelihoods are drawn from certain probability distributions, there can exist critical fractions of node failures for which the failure will cascade to the entire network. This holds when multiple networks are coupled together \cite{buldyrev2010}. 

Network-based approaches have been widely used to model the resilience of infrastructure \cite{zimmerman2016,gao2016,derrible2017}. This is in addition to conceptual frameworks \cite{park2013,linkov2014,bruneau2003}, highly detailed hazard simulations \cite{han2009, staid2014,ouyang2012,booker2010}, and statistical and machine learning approaches \cite{nateghi2011,nateghi2018,mukherjee2018,arab2015,shashaani2018} \footnote{See \citep{ouyang2014} for a comprehensive list of topics.}. All of this work contributes greatly toward improving the resilience of infrastructure by advancing theoretical understandings in networks science \cite{gao2016}, addressing particular infrastructure inefficiencies \cite{fang2017}, and improving policy decisions \cite{guikema2018}. 

Generalized graph-theoretic resilience analyses commonly model disruptions by assigning a probability of failure to each vertex in the graph \cite{gao2016,hu2016,buldyrev2010}. The random pattern of outages fits within a probabilistic formalism allowing for a theoretical understanding of network properties, but provides little realism in the spatial pattern of disruptions. Many of the infrastructure systems analyses continue to use random vertex failures as the general form of the disruption \cite{erdener2014,praks2015,larocca}. Degree targeting is another commonly used technique in which failures are initiated at vertices with the highest degree \cite{hines2010,duenas-osorioleonardo2007,winklerjames2011,hu2016,duenas-osorio2009}. This method is representative of a targeted attack in which an agent wishes to remove nodes which connect to a large portion of the network, however, there is no restriction on the spatial distribution of the failures. Similarly, other vertex properties have been used to motivate targeting such as betweenness \cite{hines2010} or maximum flow \cite{duenas-osorio2009}. Localized failures---in which failures are initialized in small connected components---have been previously studied, however with limited scope; focusing primarily on repair strategies \cite{hu2016}, or to replicate previous incidents \cite{praks2015}. 

In this work, we isolate the importance of accounting for the spatial distribution of a disruption and show that inducing changes in \textit{only} the spatial distribution significantly impacts measurements of system performance. Specifically, the goal of the analysis is not so much to propose a particular spatial pattern of disruption over another, but to demonstrate the importance of considering the shape of disruptions in estimating infrastructure recovery. We present the results in a case study of an electric power distribution grid's response to a hurricane. The electric power distribution system has been identified as a critical component of assessing the vulnerability of the electric power grid to severe-weather disruptions such as hurricanes, with approximately 90\% of outages occurring at the distribution level \cite{ji2016}. 



\section*{Methods}
\begin{figure}[!h]
\centering
\includegraphics[width=\textwidth]{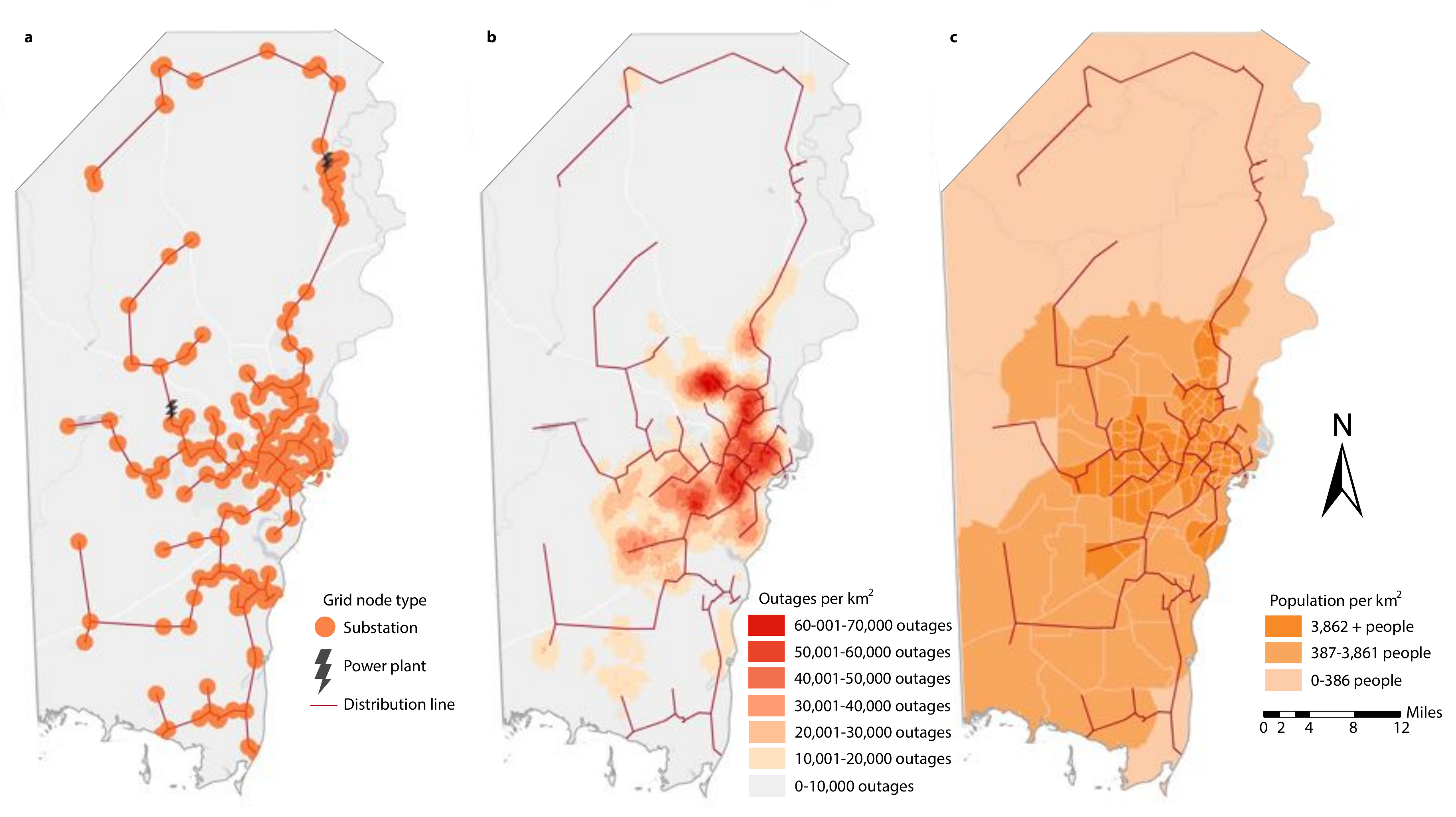}
\caption{\label{fig:pwrNetwork} The case study network situated in the Gulf Coast of the U.S. \textbf{a} The layout of the electric power grid placed over the county. \textbf{b} The density of customer-level power outages during Hurricane Katrina with the network overlain. \textbf{c} Census-tract level population density for the corresponding area.} 
\end{figure}

As mentioned earlier, to investigate the sensitivity of infrastructure system performance to the spatial distribution of disruptions, we present the case of an electric power distribution system's recovery after a major landfalling hurricane. Specifically, we focus on the impact of the \emph{spatial} distribution of hurricane-induced disruptions on the performance of an electric power grid located in the Gulf Coast of the U.S. (Figure \ref{fig:pwrNetwork}). We do this by simulating large-scale disruptions in the distribution grid, mapping the hurricane-induced disruptions to component failures (outages) in a distribution-level power grid and studying the sensitivity of the resilience of the system to the spatial distribution of the disruption. The simulated outages are subsequently repaired over time, replicating the actual recovery of the power grid from the hurricane disruption so as to study the dynamics of the system's recovery. 

\subsection*{Electric Power Network}\label{power}

The city for which this analysis is being performed provided GIS files including the location of all of the county's power substations. These are used to locate the position of the nodes in the test network. There are 221 substations and 2 power plants in this data. As we were unable to retrieve information on the connections between the substations,  nodes are connected using a minimum spanning tree to establish the edges of the graph. The resulting graph has 223 vertices and 222 edges.

\subsection*{Disruption generation algorithms}
In this section, we describe the different disruption patterns evaluated in this study. All cases described cause failures in 60\% of the vertices, and this failure proportion is kept constant through all trials. This is in accordance with the actual impact of Hurricane Katrina on the electric power distribution network under study. As previous work primarily focuses on analyzing randomized failures, we use random outages as a base for comparison with previous studies. In simulation replication, a different set of vertices is chosen at random such that 60\% of the network is inoperable. The random disruptions form a \emph{control sample} as there is explicitly no spatial association among the initial disruption. 

To evaluate how the spatial characteristics of the disruption impact the network, additional simulation trials are performed using disruptions generated by search trees. Disruptions are generated using both a Breadth-First search (BFS) and a Depth-First search (DFS) tree \cite{bondy2008} as both create spatially constrained patterns of outages while using no intrinsic information about the individual vertices. Details of the algorithms used to generate the disruptions are listed in Algorithms \ref{alg:bfs} and \ref{alg:dfs}.

\begin{algorithm}
\caption{Breadth-First Search}\label{alg:bfs}
\begin{algorithmic}[1]
\Procedure{BFS}{graph = $G$, root = $r$, size = $n$}
\State $\textit{Q} \gets \text{empty list of vertices to search}$
\State $\textit{T} \gets \text{empty list of vertices in the tree}$
\State append $r$ to $Q$
\While {$|T|<n$} 
    \State{consider $v$, the first element of $Q$}
    \State{remove $v$ from $Q$}
    \State{append $v$ to $T$}
        \ForAll{$w$ in \textit{neighbors}(v)}
            \If{$w$ is not in $T$} 
                \State{append $w$ to $Q$}
            \EndIf
        \EndFor
\EndWhile
\EndProcedure
\Return{T}
\end{algorithmic}
\end{algorithm}

\begin{algorithm}
\caption{Depth-First Search}\label{alg:dfs}
\begin{algorithmic}[1]
\Procedure{DFS}{graph = $G$, root = $r$, size = $n$}
\State $\textit{Q} \gets \text{empty list of vertices to search}$
\State $\textit{T} \gets \text{empty list of vertices in the tree}$
\State append $r$ to $Q$
\While {$|T|<n$} 
    \State{consider $v$, the first element of $Q$}
    \State{remove $v$ from $Q$}
    \State{append $v$ to $T$}
    \If{$w \in \textit{neighbors}(v), w \not\in T $}
        \State{append $w$ to front of $Q$}
    \EndIf
\EndWhile
\EndProcedure
\Return{T}
\end{algorithmic}
\end{algorithm}

 A BFS begins at a random vertex in the network and failures propagate to all neighbors of that vertex before extending to neighbors-of-neighbors. This provides a method for generating localized clusters of failures. Similarly, a DFS outage pattern begins at a random vertex and progresses away from the root node to a maximal length before searching additional root-node neighbors. The spatial pattern of DFS trees are connected, but far less localized. These are referred to as the the BFS and DFS disruption methods for the remainder of the paper. 
 
\begin{figure}[!h]
\centering
\includegraphics[width=\textwidth]{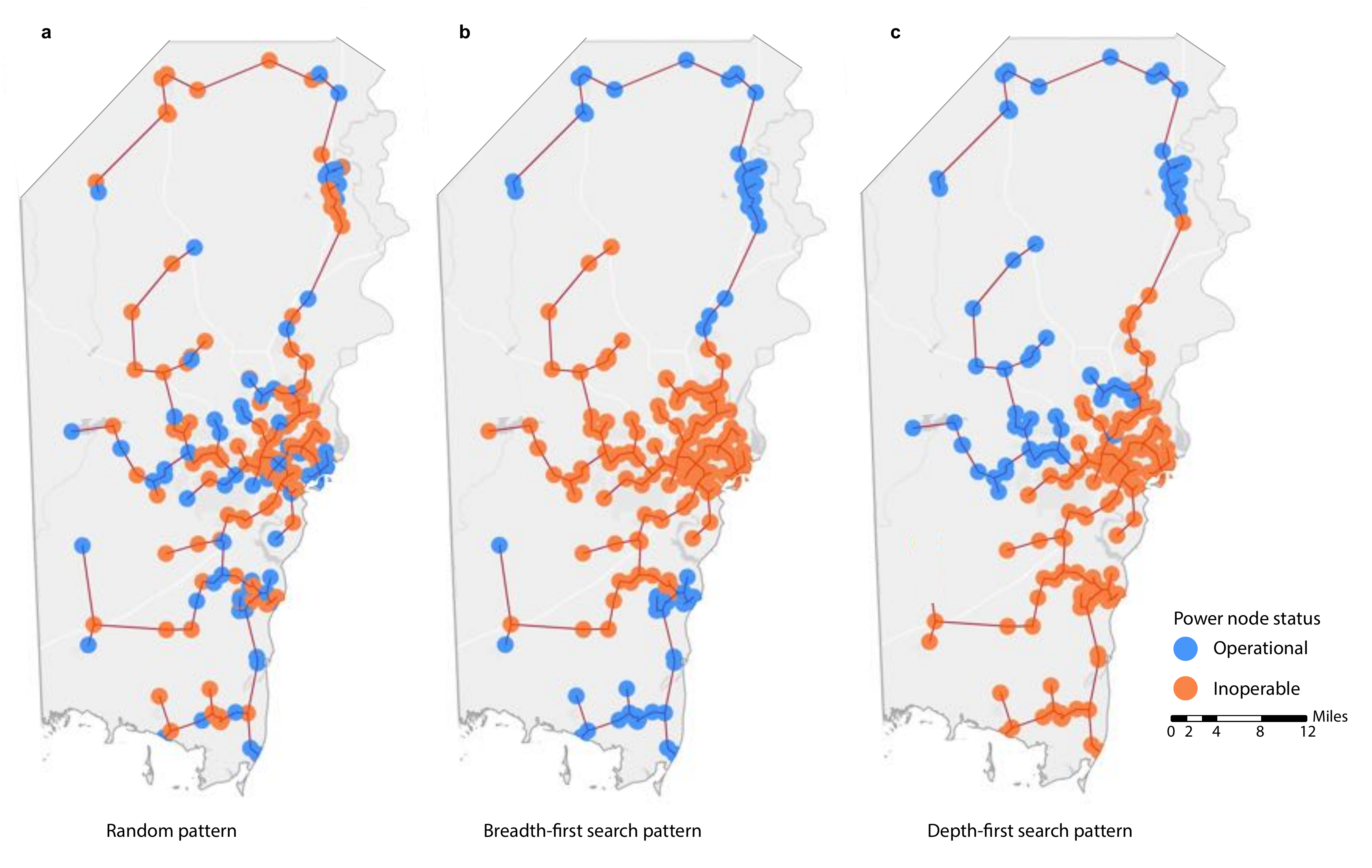}
\caption{\label{fig:threetypes}Outage generation types. The results of three outage generation techniques, each inducing failures in 60\% of the grid. Figure \textbf{a} is one instance of an outage generated randomly. Figure \textbf{b} is a an outage generated using a breadth-first algorithm, while \textbf{c} is a depth-first algorithm. }
\end{figure}

 The search tree generation methods are computationally cheap, and are built entirely using the spatial structure of the network. The selection of these algorithms are motivated by existing research supporting the existence of tree-shaped outages in distribution systems owing to the hierarchical nature of electric power distribution \cite{ji2016,dobson2016}. Here, we do not validate actual spatial distributions of outages against the BFS and DFS generation methods, but instead use these methods to isolate the significance of different spatial configurations of outages in the network on measurements of system performance. The initial distribution of outages for one simulation replication are seen in Figure \ref{fig:threetypes}.  

\subsection*{Simulation Methodology}

The recovery simulation  generates initial disruptions via random, BFS and DFS methods then subsequently repairs vertices in the network. The rate of repair (i.e., repaired vertices per time unit) is derived from the rate of outages seen in the gulf-coast power operator data. This rate is kept constant through all experiments. At every time step, the vertices to be repaired are chosen based on their contribution to the total network efficiency. The number of vertices to be repaired is first fixed based on the time dependent repair rate, then the set of vertices chosen for repair are selected from the subset of inoperable vertices which---if repaired---would maximally improve the network efficiency. Vertices are selected in a greedy fashion such that the selected subset maximally improves the efficiency of the network. The heuristic search is detailed in Algorithm \ref{alg:opt}. 

\begin{algorithm}
\caption{Local-optimal search. \\
\small{Here, \textit{GE} is the global efficiency of a graph, and $F-R$ indicates the removal of vertices R from F.}}\label{alg:opt}
\begin{algorithmic}[1]
\Procedure{LocalOpt}{graph = $G$, failed vertices = $F$,repair= $n$}
\State $\textit{R} \gets \text{empty list of vertices to be repaired}$
    \If{ $|V(F)|=|V(G)|$}
        \State{$R=\text{vertex with maximum degree}$}
        \State{$F=F -R$}
        \State{LocalOpt(G,F,n-1)}
    \Else{ $|V(F)|<|V(G)|$}
        \If{$|V(F)|+n \geq |V(G)|$}
            \State{$R=F$}
        \Else{$|V(F)|+n < |V(G)|$}
            \State{$R=f \in F \quad \text{s/t} \; \textit{GE}(G+f)\geq \textit{GE}(G+f') \; \forall f' \in F\,  \text{and}\, f' \not=f$}
        \EndIf  
    \EndIf
\EndProcedure
\Return{T}
\end{algorithmic}
\end{algorithm}

\begin{figure}[h]
\centering
\includegraphics[width=\textwidth]{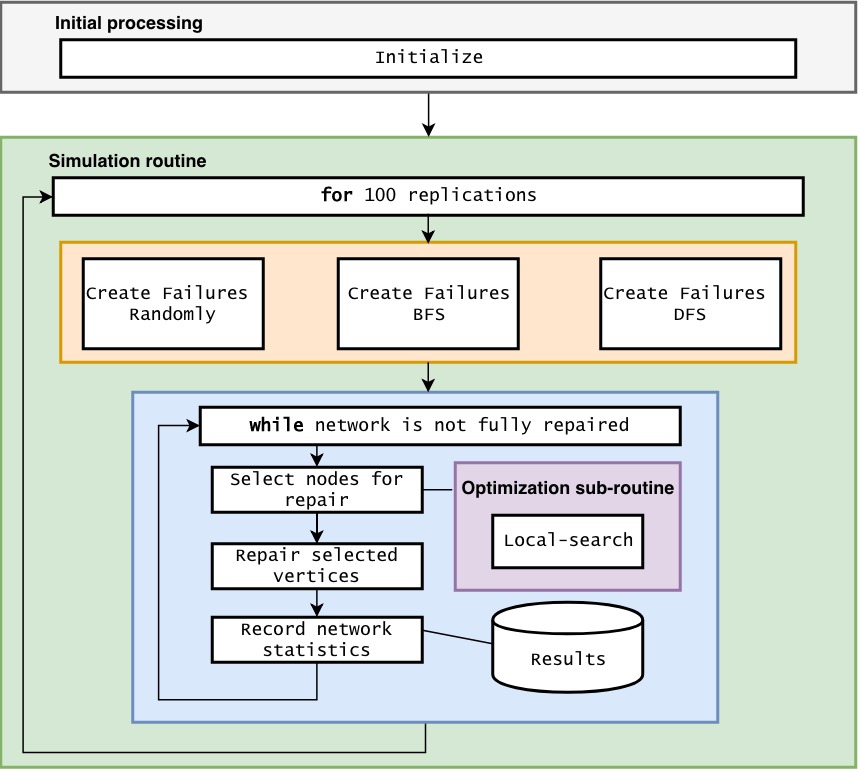}
\caption{\label{fig:simflow} An overview of simulation methodology. The process here represents one simulation iteration. }
\end{figure}

Network statistics are recorded at each step and vertices are repaired until the network is fully operational. The simulation procedure is depicted in Figure \ref{fig:simflow}. The process of creating disruptions and repairing is repeated 100 times for each disruption generation method to account for the inherent randomness in the generation of the initial distributions. The analyses were performed on a 16-core Intel Xeon W-2145 processor, each operating at 3.7GHz with 32GB of ram. Simulation, analysis, and resulting plots were all generated in R version 3.4.4 \cite{rbase}. Network statistics were calculated using \verb+ igraph+ \cite{igraph}. 

\subsection*{Performance metric calculation}
We measure the \emph{global efficiency} of the electric power network as it fails and recovers as one dimension of network performance. Global efficiency is defined as \\
\[
\text{Eff}(G)=\frac{1}{n(n-1)}\sum_{i<j \in G} \frac{1}{d(i,j)}
\]
\noindent
where $d(i,j)$ is the distance between vertex pair $i$ and $j$. Network efficiency as a concept was proposed as a measure of how efficiently a network exchanges information \cite{latora2001}. It has been evaluated  in the context of power system resilience evaluation \cite{larocca} and used as a proxy for network performance \cite{sun2017,winklerjames2011}.

Additionally we measure the size of the largest connected component (LCC). This is defined as the number of vertices in the largest connected subgraph \cite{newman2018}. A connected subgraph is a subset of the vertices and edges for which a path exists between all pairs of vertices. LCC has previously been used to evaluate topological models \cite{larocca} and provides a measure of the connectedness of the network (\emph{ie} a fully connected network has a maximal LCC because every vertex is included in the largest cluster).

\section{Results}

\subsection*{Static measures of impact}
\begin{figure}[!h]
\centering
\includegraphics[width=\textwidth]{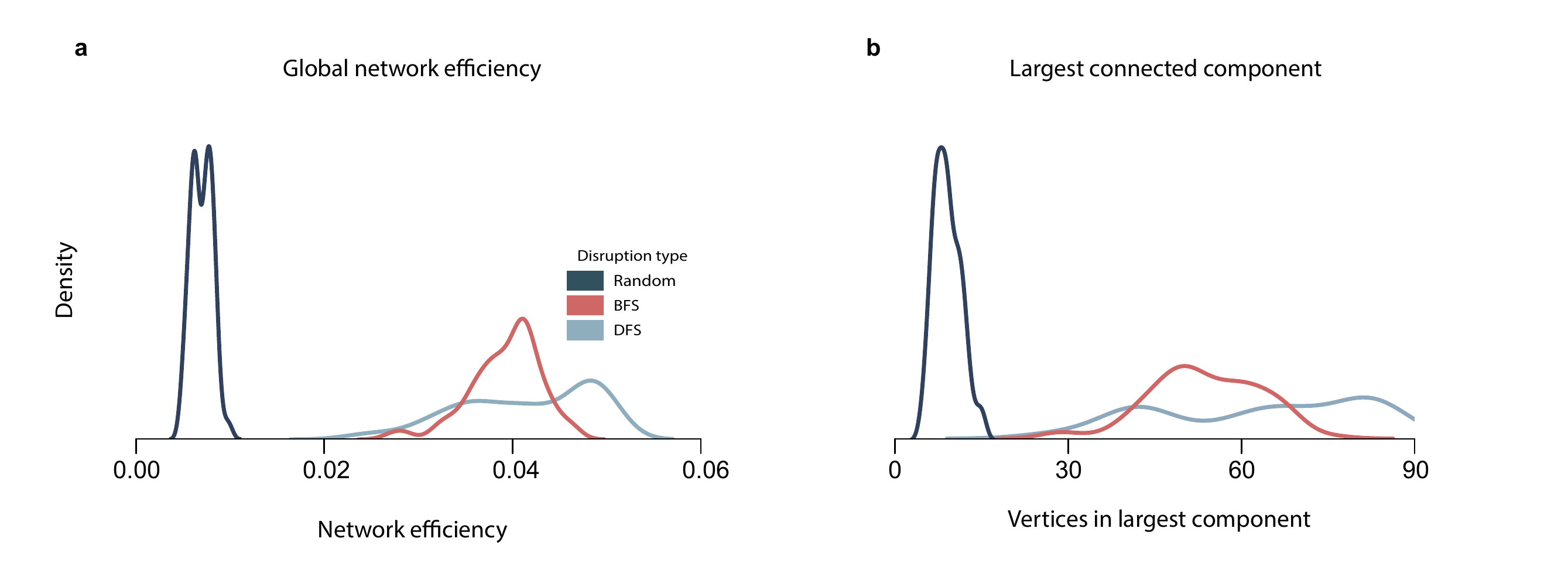}
\caption{\label{fig:geHistogram}Static disruption comparison. Relative density of network performance after 100 disruptions for each disruption generation method.  \textbf{a} is the network efficiency for all three disruption generation methods while \textbf{b} is the size of the largest connected component.}
\end{figure}

We first evaluate the sensitivity of the \textit{static measure of performance}---i.e., the performance of the system at the moment the disruption occurs---to the spatial distribution of the disruption generated randomly as well as via BFS and DFS algorithms (Figure \ref{fig:geHistogram}). To provide an equal comparison---and in accordance to real data from Hurricane Katrina---we present results which impact 60\% of the network regardless of the method of outage generation. However, our extensive sensitivity analysis suggested that the results remained consistent when evaluating network failures ranging from 10\% to 90\%. 

Computed for 100 stochastic disruptions of each type, there is significant evidence that the disruption methods alter the resilience of the system. The mean efficiency of BFS- and DFS-constructed disruptions are 485\% and 457\% higher than randomly constructed disruptions respectively. 
Mean values vary significantly at each failure size as seen in Table \ref{tab:GEsummary}. Mean LCC increases similarly with BFS disruptions---BFS increase of 595\% over random, DFS increase of 494\% over random (Table \ref{tab:LCCsummary}). Results additionally indicate sample variance increases for tree-constructed disruptions in both performance metrics as seen in Tables \ref{tab:GEsummary} and \ref{tab:LCCsummary}. In the case of the mean comparison, the distributions of efficiency and LCC values are compared using Kolmorogov-Smirnov (KS) two-sample tests and all comparisons are found to be statistically significant at a significance level of 0.01. Results of the KS tests are seen in Table \ref{tab:statGE}.

\begin{table}[!h]

\centering
\caption{\label{tab:GEsummary} Summary statistics for the distribution of efficiency for respective failure modes. Failure fraction represents the fraction of the network which was induced as failed in each iteration. Results presented here are for failures in 60\% of the network. Complete results are presented in Appendix Table \ref{tab:GEsummaryA}.}

\begin{tabular}{@{}p{1.8cm}lp{1.8cm}lll@{}}
\toprule
Generation method & Mean   & Standard deviation & Median & Min    & Max    \\ \midrule

Random                        & 0.0070 & 0.0011             & 0.0070 & 0.0047 & 0.0100 \\
BFS                           & 0.0414 & 0.0071             & 0.0420 & 0.0240 & 0.0494 \\
DFS                           & 0.0393 & 0.0038             & 0.0401 & 0.0270 & 0.0463 \\  \bottomrule
\end{tabular}\\

\end{table}

The lower efficiency values and LCC of the random disruption method indicate greater disruption in the system. Consequently any claim resulting from a measure of resilience is sensitive to the spatial characteristics of the initial disruption. Likewise, accounting for the spatial distribution of disruptions introduces greater uncertainty into our estimation of the resilience of a system. 

\begin{table}[!h]
\begin{centering}
\caption{\label{tab:statGE}P-values for two-sample, two tailed, Kolmogorov-Smirnov tests between the efficiency and LCC of given initial failure methods and failure fraction. Results at the 0.6 failure fraction are presented in this article. Results use a significance level of $\alpha=0.05$. Values of zero listed with one significant digit indicate $p< 1.11022 \text{e}-16$; this cutoff is the numerical precision of the machine used for computations. }

\begin{tabular}{lp{1.8cm}p{1.5cm}p{1.1cm}p{1.5cm}p{1.5cm}p{1.3cm}}
                 & \multicolumn{3}{c}{Efficiency}                                  & \multicolumn{3}{c}{LCC}                    \\ \hline
\multicolumn{1}{p{1.5cm}|}{Failure fraction} & Random vs BFS & Random vs DFS & \multicolumn{1}{l|}{BFS vs DFS} & Random vs BFS & Random vs DFS & BFS vs DFS \\ \hline
\multicolumn{1}{l|}{0.1}              & 0             & 0             & \multicolumn{1}{l|}{0.0039}     & 0             & 0             & 0.0541     \\
\multicolumn{1}{l|}{0.2}              & 0             & 0             & \multicolumn{1}{l|}{0.0004}     & 0             & 0             & 0.0001     \\
\multicolumn{1}{l|}{0.3}              & 0             & 0             & \multicolumn{1}{l|}{0.0014}     & 0             & 0             & 0.0000     \\
\multicolumn{1}{l|}{0.4}              & 0             & 0             & \multicolumn{1}{l|}{0.0014}     & 0             & 0             & 0.0000     \\
\multicolumn{1}{l|}{0.5}              & 0             & 0             & \multicolumn{1}{l|}{0.0001}     & 0             & 0             & 0.0000     \\
\multicolumn{1}{l|}{0.6}              & 0             & 0             & \multicolumn{1}{l|}{0.0000}     & 0             & 0             & 0.0000     \\
\multicolumn{1}{l|}{0.7}              & 0             & 0             & \multicolumn{1}{l|}{0.0000}     & 0             & 0             & 0.0008     \\
\multicolumn{1}{l|}{0.8}              & 0             & 0             & \multicolumn{1}{l|}{0.0000}     & 0             & 0             & 0.0000     \\
\multicolumn{1}{l|}{0.9}              & 0             & 0             & \multicolumn{1}{l|}{0.0000}     & 0             & 0             & 0.0000    \\
\end{tabular}
\end{centering}

\end{table}

\begin{table}[!h]

\centering
\caption{\label{tab:LCCsummary} Summary statistics for the distribution of largest connected component (LCC) for respective failure modes. Results presented here are for failures in 60\% of the network. Complete results are presented in Appendix Table \ref{tab:LCCsummarya}.}
\begin{tabular}{@{}p{1.8cm}lp{1.8cm}lll@{}}
\toprule
Generation method  & Mean   & Standard deviation & Median & Min   & Max    \\ \midrule
Random                       & 9.05   & 2.32               & 9.00   & 5.00  & 15.00  \\
BFS                    & 62.94  & 17.71              & 66.50  & 28.00 & 83.00  \\
DFS                   & 53.75  & 9.54               & 53.00  & 28.00 & 76.00  \\ \bottomrule
\end{tabular}
\end{table}

The sensitivity of the resilience to disruption method additionally manifests when measuring the number of customers with restored power. Mapping the geographical location of each of the vertices in our network to their respective census tract allows us to allocate customers to each substation relative to their population density. Using this this approximation, an average of 40.60\% of the customers retain power when disrupted randomly, versus 39.21\% and 39.47\% for BFS and DFS outages respectively. This similarity is expected as the disruptions are constructed to disconnect 60\% of the substations in the network, leaving approximately 40\% of the network operational. However similar to measurements of efficiency and LCC, the variance among population affected is higher for tree-based disruptions. Table \ref{tab:custStat} shows the distribution of the number of customers without power after the network is made inoperable. After random outages are induced in the system 33.57\%--48.35\% of the population's distribution level power remains operational, while after BFS and DFS outages 26.54\%--53.77\% and 26.94\%--48.95\% of the population's power remain operational respectively. This represents an 88\% increase in the uncertainty of the performance estimates. Providing estimates of uncertainty is critical to decision makers for the accurate characterization of the resilience of a system \cite{flage2014}.

\begin{table}[tbh]

\centering
\caption{\label{tab:custStat}Summary statistics for the distribution of percent of county customers without power in a static analysis. All numbers represent the fraction of the total population of the county without power.}
\begin{tabular}{@{}lccccc@{}}
\toprule
             & Mean    & Std Dev & Median  & Min     & Max    \\ \midrule
Random       & 0.5928  & 0.0351 & 0.5940 & 0.5165 & 0.6643  \\
BFS          & 0.5909  &  0.0676& 0.6079 & 0.4623  & 0.7346  \\
DFS          & 0.6151  & 0.0651 &  0.6053  & 0.5105&   0.7306 \\  \bottomrule
\end{tabular}

\end{table}

\subsection*{Dynamic measures of impact}

\begin{figure}[!h]
\centering
\includegraphics[width=\textwidth]{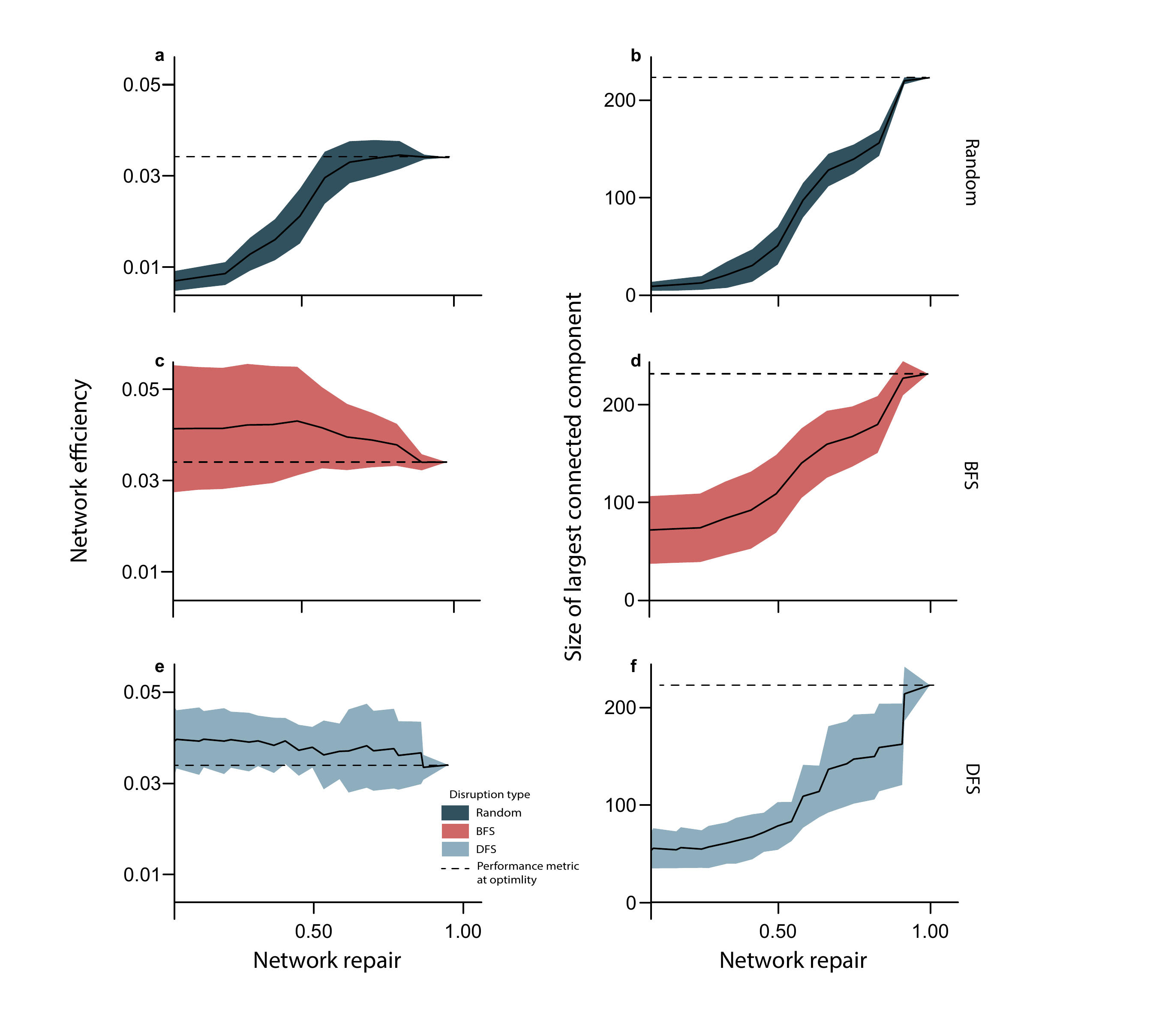}
\caption{\label{fig:geDynamics}Performance metrics measured after the disruption over time for each disruption method. In \textbf{a-f}, the bands of uncertainty represent 95\% confidence intervals sampled from the empirical density at each point in time. The black line is the mean of the observations. The x-axis is the relative-completeness of the network repair scaled by the total restoration time for each replication.} 
\end{figure}

We also evaluate the \textit{dynamic performance} ---i.e., time dependant performance metrics---under separate initial disruption methods as the power grid is repaired (Figure \ref{fig:geDynamics}). The system performance---characterized by efficiency and LCC---is then measured over time as the system recovers. This is done to characterize the dynamic resilience of the grid under each disruption generation method, ceteris paribus. 

Despite holding the recovery process constant, these results show the efficiency of the network differs greatly in overall functional form between random and spatially generated disruptions, indicating the recovery is significantly coupled to the spatial distribution of disruptions. Recovery from a random disruption pattern increases over time, reaching a maximum prior to all nodes being repaired (Figure \ref{fig:geDynamics}e). This is an indication of the network exhibiting \emph{antifragile} properties in which a full reconstruction of the network is not optimal with respect to the chosen performance metric \cite{aven2015,taleb2012,fang2017}. Spatially-constructed outages generally have a much higher efficiency throughout but follow an entirely different functional form than the recovery from random disruptions. The deviation between mean efficiency is highest at the initial disruption and decreases over time. Similar to the static analysis, the variance is larger in the recovery from spatially characterized outages. Thus, failing to account for the spatial characteristics of the network disruption can drastically change implications drawn from the associated resilience analysis. A key difference is the lack of antifragility in the distribution electric power network with spatially characterized outages.

The difference between the disruption generation techniques is diminished when comparing the dynamics of the mean LCC rather than mean network efficiency (Figure \ref{fig:geDynamics} b,d,f). Beyond the initial value of the LCC at the time of failure, there is little difference in the functional form of the recovery of the network. The size of the LCC in the network generally increases at an increasing rate when vertices are repaired in the network, the primary difference being the initial size of the LCC after failures are generated in the network. These estimates of system recovery are therefore dependant on the spatial characteristics of the initial disruption; however, this result is sensitive to the performance metric used to measure recovery.

\section{Conclusion}
A key element of resilience is the ability of a system to respond to and recover from disruptions of unprecedented magnitude or unforeseen cause. By their nature, \emph{all} disruptions will require recovery. This positions system recovery as a critical measurement in evaluating the multifaceted resilience of infrastructure systems. A holistic understanding of all types of community recovery is imperative for the continued adaptation to unforeseen challenges. However, these holistic understandings must be built upon a foundational knowledge of the interaction of disasters with the built environment. We contribute to the knowledge related to the interaction of the power distribution grid and hurricanes by providing a novel framework for network resilience analysis which is agnostic to the specifics of the system, allowing for general insights about all facets of community recovery. Our framework for considering spatially-constrained disruptions can be applied to any hierarchical network within a community adversely effected by natural hazards. 

We show that the post-disruption performance of the electrical power distribution grid is highly sensitive to the spatial characteristics of disruptions in the system. Consequently, any insights about general grid resilience which fail to account for the spatial characteristics of the hazard significantly misrepresent the impact of natural hazards on distribution-level electric power infrastructure. More specifically, through the repeated simulation of multiple methods of failure and recovery, we show that previous methods of evaluating disaster impact overestimate the certainty associated with the measurements of system recovery. We show via multiple avenues that improved characterizations of disaster impact significantly increase \emph{both} the magnitude and uncertainty of the initial impact in the system. This difference holds through the duration of the recovery process; and when considering the dynamics of the system we find that emergent system properties such as antifragility are also dependant on the characteristics of the initial disruption. These differences are most striking when contextualized by their impact on the power distribution grid at a customer level. Our estimates indicate that the estimated range of customers with access to electricity varies from 33-48\% of the county using previous methods, and up to 26-53\% when using improved outage characterizations, highlighting the need for continued study of both the pattern of impacts due to natural disasters and the vulnerability of the electric power distribution grid. By demonstrating the sensitivity of the spatial distribution of outages on the electric power grid, we hope to encourage  consideration of the spatial distribution of disruptions in conducting infrastructure resilience analytics.

\section*{Appendix}
\setcounter{table}{0}
\renewcommand*\thetable{A\arabic{table}}

\begin{table}[H]

\centering
\caption{\label{tab:GEsummaryA} Summary statistics for the distribution of efficiency for respective failure modes. Failure fraction represents the fraction of the network which was induced as failed in each iteration. Results presented in the body of the work represent a failure fraction of 0.6}

\begin{tabular}{@{}p{1.8cm}p{1.5cm}lp{1.8cm}lll@{}}
\toprule
Generation method & Failure fraction & Mean   & Standard deviation & Median & Min    & Max    \\ \midrule
Random            & 0.1              & 0.0178 & 0.0025             & 0.0174 & 0.0138 & 0.0245 \\
BFS               & 0.1              & 0.0298 & 0.0046             & 0.0297 & 0.0215 & 0.0363 \\
DFS               & 0.1              & 0.0308 & 0.0042             & 0.0320 & 0.0217 & 0.0363 \\ \midrule
Random            & 0.2              & 0.0124 & 0.0016             & 0.0121 & 0.0095 & 0.0191 \\
BFS               & 0.2              & 0.0294 & 0.0050             & 0.0278 & 0.0213 & 0.0408 \\
DFS               & 0.2              & 0.0312 & 0.0066             & 0.0293 & 0.0216 & 0.0410 \\ \midrule
Random            & 0.3              & 0.0099 & 0.0013             & 0.0097 & 0.0076 & 0.0134 \\
BFS               & 0.3              & 0.0307 & 0.0056             & 0.0295 & 0.0236 & 0.0427 \\
DFS               & 0.3              & 0.0324 & 0.0075             & 0.0286 & 0.0232 & 0.0443 \\ \midrule
Random            & 0.4              & 0.0084 & 0.0010             & 0.0083 & 0.0059 & 0.0119 \\
BFS               & 0.4              & 0.0313 & 0.0043             & 0.0294 & 0.0257 & 0.0410 \\
DFS               & 0.4              & 0.0310 & 0.0033             & 0.0304 & 0.0243 & 0.0390 \\ \midrule
Random            & 0.5              & 0.0076 & 0.0009             & 0.0075 & 0.0058 & 0.0099 \\
BFS               & 0.5              & 0.0360 & 0.0055             & 0.0358 & 0.0255 & 0.0467 \\
DFS               & 0.5              & 0.0340 & 0.0023             & 0.0343 & 0.0279 & 0.0384 \\ \midrule
Random            & 0.6              & 0.0070 & 0.0011             & 0.0070 & 0.0047 & 0.0100 \\
BFS               & 0.6              & 0.0414 & 0.0071             & 0.0420 & 0.0240 & 0.0494 \\
DFS               & 0.6              & 0.0393 & 0.0038             & 0.0401 & 0.0270 & 0.0463 \\ \midrule
Random            & 0.7              & 0.0068 & 0.0015             & 0.0069 & 0.0043 & 0.0119 \\
BFS               & 0.7              & 0.0479 & 0.0127             & 0.0462 & 0.0264 & 0.0751 \\
DFS               & 0.7              & 0.0506 & 0.0052             & 0.0508 & 0.0322 & 0.0659 \\ \midrule
Random            & 0.8              & 0.0071 & 0.0016             & 0.0069 & 0.0035 & 0.0126 \\
BFS               & 0.8              & 0.0561 & 0.0174             & 0.0533 & 0.0294 & 0.0880 \\
DFS               & 0.8              & 0.0699 & 0.0126             & 0.0724 & 0.0427 & 0.0867 \\ \midrule
Random            & 0.9              & 0.0097 & 0.0031             & 0.0097 & 0.0045 & 0.0184 \\
BFS               & 0.9              & 0.0659 & 0.0208             & 0.0609 & 0.0392 & 0.1389 \\
DFS               & 0.9              & 0.0971 & 0.0140             & 0.0961 & 0.0600 & 0.1372 \\ \bottomrule
\end{tabular}\\

\end{table}

\begin{table}[H]

\centering
\caption{\label{tab:LCCsummarya} Summary statistics for the distribution of largest connected component (LCC) for respective failure modes.}
\begin{tabular}{@{}p{1.8cm}p{1.5cm}lp{1.8cm}lll@{}}
\toprule
Generation method & Failure fraction & Mean   & Standard deviation & Median & Min   & Max    \\ \midrule
Random            & 0.1              & 84.46  & 23.86              & 83.00  & 42.00 & 138.00 \\
BFS               & 0.1              & 158.22 & 35.67              & 160.50 & 82.00 & 202.00 \\
DFS               & 0.1              & 164.34 & 35.36              & 181.00 & 80.00 & 202.00 \\ \midrule
Random            & 0.2              & 41.89  & 12.49              & 40.00  & 20.00 & 87.00  \\
BFS               & 0.2              & 122.21 & 32.35              & 125.00 & 82.00 & 179.00 \\
DFS               & 0.2              & 125.23 & 43.74              & 111.50 & 66.00 & 180.00 \\ \midrule
Random            & 0.3              & 25.54  & 8.05               & 24.00  & 13.00 & 50.00  \\
BFS               & 0.3              & 105.79 & 29.70              & 91.00  & 64.00 & 154.00 \\
DFS               & 0.3              & 105.84 & 40.11              & 80.50  & 54.00 & 159.00 \\ \midrule
Random            & 0.4              & 16.90  & 4.33               & 16.00  & 9.00  & 29.00  \\
BFS               & 0.4              & 85.47  & 19.73              & 82.00  & 44.00 & 129.00 \\
DFS               & 0.4              & 73.91  & 13.19              & 70.00  & 44.00 & 124.00 \\ \midrule
Random            & 0.5              & 12.22  & 2.83               & 12.00  & 7.00  & 21.00  \\
BFS               & 0.5              & 76.04  & 17.13              & 82.00  & 39.00 & 100.00 \\
DFS               & 0.5              & 62.14  & 7.92               & 64.00  & 41.00 & 80.00  \\ \midrule
Random            & 0.6              & 9.05   & 2.32               & 9.00   & 5.00  & 15.00  \\
BFS               & 0.6              & 62.94  & 17.71              & 66.50  & 28.00 & 83.00  \\
DFS               & 0.6              & 53.75  & 9.54               & 53.00  & 28.00 & 76.00  \\ \midrule
Random            & 0.7              & 6.80   & 1.51               & 7.00   & 4.00  & 12.00  \\
BFS               & 0.7              & 46.78  & 15.55              & 47.00  & 17.00 & 68.00  \\
DFS               & 0.7              & 47.43  & 7.66               & 49.00  & 24.00 & 66.00  \\ \midrule
Random            & 0.8              & 5.03   & 1.01               & 5.00   & 3.00  & 8.00   \\
BFS               & 0.8              & 28.57  & 9.98               & 28.00  & 10.00 & 46.00  \\
DFS               & 0.8              & 35.42  & 9.02               & 38.00  & 16.00 & 46.00  \\ \midrule
Random            & 0.9              & 3.40   & 0.57               & 3.00   & 3.00  & 5.00   \\
BFS               & 0.9              & 11.49  & 4.49               & 10.00  & 5.00  & 24.00  \\
DFS               & 0.9              & 16.25  & 3.35               & 16.00  & 9.00  & 24.00  \\ \bottomrule
\end{tabular}
\end{table}

%

\newpage
\nolinenumbers
\section{References}

\bibliographystyle{model1-num-names}
\bibliography{prelim.bib}

\end{document}